\documentclass[12pt,a4paper,final]{revtex4}

\usepackage{sidecap}
\usepackage{ulem}
\usepackage{epsfig}
\usepackage{amsmath,amssymb,amsthm}
\usepackage{graphicx}
\usepackage{bm}
\usepackage{color}

\usepackage{enumerate}

\usepackage{float}
\usepackage{tabularx}

\DeclareMathAlphabet{\mathpzc}{OT1}{pzc}{m}{it}

\begin{document}

\renewcommand{\textfraction}{0.00}


\newcommand{\vAi}{{\cal A}_{i_1\cdots i_n}} 
\newcommand{\vAim}{{\cal A}_{i_1\cdots i_{n-1}}} 
\newcommand{\vAbi}{\bar{\cal A}^{i_1\cdots i_n}}
\newcommand{\vAbim}{\bar{\cal A}^{i_1\cdots i_{n-1}}}
\newcommand{\htS}{\hat{S}} 
\newcommand{\htR}{\hat{R}}
\newcommand{\htB}{\hat{B}} 
\newcommand{\htD}{\hat{D}}
\newcommand{\htV}{\hat{V}} 
\newcommand{\cT}{{\cal T}} 
\newcommand{\cM}{{\cal M}} 
\newcommand{\cMs}{{\cal M}^*}
\newcommand{\vk}{\vec{\mathbf{k}}}
\newcommand{\bk}{\bm{k}}
\newcommand{\kt}{\bm{k}_\perp}
\newcommand{\kp}{k_\perp}
\newcommand{\km}{k_\mathrm{max}}
\newcommand{\vl}{\vec{\mathbf{l}}}
\newcommand{\bl}{\bm{l}}
\newcommand{\bK}{\bm{K}} 
\newcommand{\bb}{\bm{b}} 
\newcommand{\qm}{q_\mathrm{max}}
\newcommand{\vp}{\vec{\mathbf{p}}}
\newcommand{\bp}{\bm{p}} 
\newcommand{\vq}{\vec{\mathbf{q}}}
\newcommand{\bq}{\bm{q}} 
\newcommand{\qt}{\bm{q}_\perp}
\newcommand{\qp}{q_\perp}
\newcommand{\bQ}{\bm{Q}}
\newcommand{\vx}{\vec{\mathbf{x}}}
\newcommand{\bx}{\bm{x}}
\newcommand{\tr}{{{\rm Tr\,}}} 
\newcommand{\bc}{\textcolor{blue}}

\newcommand{\beq}{\begin{equation}}
\newcommand{\eeq}[1]{\label{#1} \end{equation}} 
\newcommand{\ee}{\end{equation}}
\newcommand{\bea}{\begin{eqnarray}} 
\newcommand{\eea}{\end{eqnarray}}
\newcommand{\beqar}{\begin{eqnarray}} 
\newcommand{\eeqar}[1]{\label{#1}\end{eqnarray}}
 
\newcommand{\half}{{\textstyle\frac{1}{2}}} 
\newcommand{\ben}{\begin{enumerate}} 
\newcommand{\een}{\end{enumerate}}
\newcommand{\bit}{\begin{itemize}} 
\newcommand{\eit}{\end{itemize}}
\newcommand{\ec}{\end{center}}
\newcommand{\bra}[1]{\langle {#1}|}
\newcommand{\ket}[1]{|{#1}\rangle}
\newcommand{\norm}[2]{\langle{#1}|{#2}\rangle}
\newcommand{\brac}[3]{\langle{#1}|{#2}|{#3}\rangle} 
\newcommand{\hilb}{{\cal H}} 
\newcommand{\pleft}{\stackrel{\leftarrow}{\partial}}
\newcommand{\pright}{\stackrel{\rightarrow}{\partial}}

\newcommand{\squeezeup}{\vspace{-2.5mm}}

\newcommand{\RomanNumeralCaps}[1]
    {\MakeUppercase{\romannumeral #1}}


\title{Exploring the initial stages in heavy-ion collisions with high-$p_{\perp}$ $R_{AA}$ and $v_2$ theory and data}

\date{\today}
 
\author{Dusan Zigic$^1$, Bojana Ilic$^1$, Marko Djordjevic$^2$ and Magdalena Djordjevic$^1$} 
 
\affiliation{$^1$Institute of Physics Belgrade, University of Belgrade, Belgrade, Serbia \\
$^2$Faculty of Biology, University of Belgrade, Belgrade, Serbia}

\begin{abstract}

Traditionally, low-$p_\perp$ sector is used to infer the features of initial stages before QGP thermalization. On the other hand, recently acquired wealth of high-$p_{\perp}$ experimental data paves the way to utilize the high-$p_{\perp}$ particles energy loss in exploring the initial stages. We here study how four different commonly considered initial-stage scenarios – which have the same temperature profile after, but differ in the 'temperature' profile before thermalization – affect predictions of high-$p_\perp$  $R_{AA}$ and $v_2$ observables.   Contrary to common expectations, we  obtain that high-$p_\perp$ $v_2$ is insensitive to the initial stages of medium evolution, being unable to discriminate between different conditions. On the other hand, $R_{AA}$ is sensitive to these conditions, however, within the current errorbars, the sensitivity is not sufficient to distinguish between different initial stages. Moreover, we also reconsider the validity of widely-used procedure of fitting the energy loss parameters, individually for different initial-stage cases, to reproduce the experimentally observed $R_{AA}$. We here find that previously reported sensitivity of $v_2$ to different initial states is mainly a consequence of the $R_{AA}$ fitting procedure, which may lead to incorrect conclusions. On the other hand, if a global property, in particular the same average temperature, is imposed to tested temperature profiles, high sensitivity of high-$p_\perp$ $v_2$ is again obtained. We however show that this sensitivity would not be a consequence of differences in initial, but rather final, stages. Consequently, the simultaneous study of high-$p_\perp$ $R_{AA}$ and $v_2$, with consistent energy loss parametrization and stringently controlled temperature profiles, is necessary to assess sensitivity of different variables to differences in initial and final stages.
\end{abstract}

\maketitle 

\section{Introduction} 

It is by now firmly confirmed that a new state of matter $-$ the quark-gluon plasma (QGP)~\citep{QGP1,QGP2}, in which quarks, antiquarks and gluons are deconfined, is formed at the Relativistic Heavy Ion Collider (RHIC) and the Large Hadron Collider (LHC). Rare high transverse momentum (high-$p_{\perp}$) particles, which are created immediately upon the collision, are sensitive to all stages of QGP evolution, and are considered to be excellent probes~\citep{probe1,probe2,probe3,probe4} of this extreme form of matter. As these probes traverse QGP, they lose energy, which is commonly assessed through high-$p_{\perp}$ angular averaged ($R_{AA}$)~\citep{ALICER,CMSR,ATLASR,ALICER1,CMSR1,CMSR2,PHENIXR,STARR} and high-$p_{\perp}$ angular differential ($v_2$)~\citep{ALICEv2,CMSv2,ATLASv2,ALICEv2D,CMSv2D} nuclear modification factors. 

Commonly, the high-$p_{\perp}$ particles are used to study the nature of jet-medium interactions, while the low-$p_{\perp}$ particles are used to infer the bulk QGP properties. Accordingly, the scarce knowledge of the features of initial stages before QGP thermalization ($\tau < \tau_0$) was mostly inferred by utilizing data from low-$p_{\perp}$ sector~\citep{Bjoern,Aad,DeNicol} ($p_{\perp} \lesssim 5$ GeV). However, since  high-$p_{\perp}$ partons effectively probe QGP properties, which in turn depend on initial stages, the idea of utilizing high-$p_{\perp}$ theory and data in exploring the initial stages emerged. This idea acquired an additional boost, since a wealth of precision high-$p_{\perp}$ $R_{AA}$~\citep{ALICER,CMSR,ATLASR,ALICER1,CMSR1,CMSR2} and $v_2$~\citep{ALICEv2,CMSv2,ATLASv2,ALICEv2D,CMSv2D} data have recently became available. Thus, the main goal of this paper is to assess to what extent and through what observables, the initial stages of QGP evolution can be restrained by exploiting the energy loss of high-$p_{\perp}$ particles in evolving medium.

While clarifying these issues is clearly intriguing, the results of current theoretical studies on this subject are either inconclusive or questionable~\citep{Andres,DAB_novo,SHEME}, as e.g., the energy loss parameters are fitted to reproduce the experimentally observed high-$p_\perp$ $R_{AA}$ data, individually for different analyzed initial stages. The energy loss parametrization should, however, clearly be a property of high-$p_\perp$ parton interactions with the medium, rather than of individual temperature profiles. Consequently, to more rigorously study this issue, one needs a high control on both the energy loss and the analyzed temperature ($T$) profiles. To achieve this, we here use our state-of-the-art dynamical energy loss formalism, embedded in Bjorken 1+1D medium evolution~\citep{Bjorken} (DREENA-B framework~\citep{DREENAB}). Bjorken 1+1D medium evolution has a major advantage for this study, as it allows to analytically introduce different evolutions before thermalization, with the same evolution after thermalization, which therefore allows to clearly isolate only the effects of different initial stages. Consequently, we will here consider the effects on high-$p_\perp$ $R_{AA}$ and $v_2$ predictions of four common initial-stage cases~\citep{SHEME}, which have the same $T$ profiles after, but differ in $T$ profiles before the thermalization.

Furthermore, we recently demonstrated that DREENA-B framework is able to accurately reproduce both high-$p_{\perp}$  $R_{AA}$ and $v_2$ data for diverse colliding systems and energies ($Pb+Pb$ at 2.76 TeV and 5.02 TeV and $Xe+Xe$ at 5.44 TeV), for both light and heavy flavors ($h^\pm$, B, D) and all available centralities, without introducing new phenomena~\citep{monopoli,monopoli1}. This is in distinction to many other formalisms, which employ more advanced medium evolution models, but contain simplified energy loss models, which have a tendency to underestimate $v_2$ relative to the experimental data, which is widely known as the $v_2$ puzzle~\citep{vP3,v2Puzzle2}. Moreover, we recently obtained that going from 1+1D Bjorken to full 3+1D hydrodynamics evolution~\cite{DREENA_A}, does not significantly change the agreement between our predictions and experimental data, strongly suggesting that, for high-$p_\perp$ data, accurate energy loss description is more important than the medium evolution. Consequently, for this study, using 1+1D Bjorken evolution has a major advantage of a tight control over the temperature profiles used to mimic different initial states, while, at the same time, providing a reasonably realistic description of the data within our model.

 The paper is organized as follows. In Section II, theoretical and computational frameworks are outlined. In Section III, we first assess the sensitivity of $R_{AA}$ and $v_2$ to the aforementioned  initial stages. We then adopt the approach of fitting initial temperature ($T_0$) to reproduce the same $R_{AA}$ in all cases, and then assess the effect of thus obtained "modified" temperature  profiles on $R_{AA}$ and $v_2$. We finally reexamine the validity of widely-used procedure~\citep{Andres,DAB_novo,SHEME} of fitting the energy loss parameters for different initial-stage cases  to reproduce the same $R_{AA}$. For all these studies, we analytically pinpoint the origin of the obtained results. Our conclusions are presented in Section IV.  

\section{Theoretical and Computational Frameworks}

To obtain the medium modified distribution of high-$p_{\perp}$ light and heavy flavor particles, the generic pQCD  convolution formula~\citep{RunA,PLF} is utilized:
\begin{eqnarray}
\frac{E_f d^3\sigma}{dp^3_f} = \frac{E_i d^3\sigma(Q)}{dp^3_i} \otimes P(E_i \rightarrow E_f)\ \otimes D(Q \rightarrow H_Q), 
\label{pQCD}
\end{eqnarray}
where indexes $f$ and $i$ refer to the final hadron ($H_Q$) and initial parton ($Q$), respectively. $\frac{E_i d^3\sigma(Q)}{dp^3_i}$ denotes the parton initial momentum distribution, calculated according to~\citep{ID}. $P(E_i \rightarrow E_f)$ presents the energy loss probability based on our  dynamical energy loss formalism (see below). $D(Q \rightarrow H_Q)$ stands for fragmentation function of parton into the hadron ($H_Q$), where for the light hadrons, D and B mesons we apply DSS~\citep{FF}, BCFY~\citep{FF1} and KLP~\citep{FF2} fragmentation functions, respectively.

The dynamical energy loss formalism~\citep{DRad,DRad1,DColl} includes several unique features in modeling jet-medium interactions:   (1) The finite size QCD medium consisting of dynamical (moving) as opposed to static scattering centers, which allows the longitudinal momentum exchange with the medium constituents. (2) The calculations within the finite temperature generalized Hard-Thermal-Loop approach~\citep{Kapusta}, so that infrared divergences are naturally regulated in a highly non-trivial manner, contrary to many models which apply tree-level (vacuum-like) propagators~\citep{BDMPS0,ASW0,GLV0,HT0}. (3) Both radiative~\citep{DRad,DRad1} and collisional~\citep{DColl} contributions are calculated within the same theoretical framework. (4) The generalization to a finite magnetic mass~\citep{Mmass}, running coupling~\citep{RunA} and beyond the soft-gluon approximation~\citep{bsga} is performed. In this paper for magnetic to electric mass ratio we assume value $\mu_M / \mu_E = 0.5$, since various non-perturbative~\citep{xb,xb2} approaches reported it to be in the range $0.4 - 0.6$. (5) The energy loss probability comprises also multigluon~\citep{MGF} and path-length~\citep{PLF} fluctuations. The path-length fluctuations are calculated according to the procedure presented in~\citep{Dainese}, and are provided in Ref.~\citep{DREENAC}. 

As outlined in Ref.~\citep{DREENAB}, the analytical expression for single gluon radiation spectrum, in evolving medium, reads:
\begin{align}~\label{radEL}
\frac{dN_{rad}}{dx d \tau} {} & = \frac{C_2(G) C_R}{\pi}  \frac{1}{x}\int{\frac{d^2{\mathbf{q}}}{\pi}} 
\frac{d^2{\mathbf{k}}}{\pi} \frac{\mu^2_E(T) - \mu^2_M(T)}{[{\mathbf{q}}^2 + \mu^2_E(T)] [{\mathbf{q}}^2 + \mu^2_M(T)]}  T \alpha_s(ET) \alpha_s\big(\frac{{\mathbf{k}}^2 + \chi(T)}{x} \big)  \nonumber \\
\times & \Big[1-\cos{\big( \frac{({\mathbf{k}} + {\mathbf{q}})^2 + \chi(T)}{xE^+} \tau \big)} \Big]
   \frac{2 ({\mathbf{k}}+{\mathbf{q}})}{({\mathbf{k}}+{\mathbf{q}})^2 + \chi(T)} \Big[\frac{{\mathbf{k}}+{\mathbf{q}}}{({\mathbf{k}}+{\mathbf{q}})^2 + \chi(T)} - \frac{{\mathbf{k}}}{{\mathbf{k}}^2 + \chi(T)} \Big] ,
\end{align}
where ${\mathbf{k}}$ and ${\mathbf{q}}$ denote transverse momenta of radiated and exchanged gluons, respectively,  $C_2(G) = 3$, $C_R=4/3$ ($C_R=3$) for quark (gluon) jet, while $\mu_E(T)$ and $\mu_M(T)$ are electric (Debye) and magnetic screening masses, respectively. Temperature dependent Debye mass~\citep{Deb} is obtained by self-consistently solving Eq. (5) from Ref.~\citep{DREENAB}. 
$\alpha_s$ is the (temperature dependent) running coupling~\citep{Runn}, $E$ is the initial jet energy, while $\chi(T) =  M^2x^2 + m^2_g(T)$, where $x$ is the longitudinal momentum fraction of the jet carried away by the emitted gluon, $M$ is the mass of the quark ($M_{u,d,s} \approx \mu_E(T)/ \sqrt{6}$ i.e., the thermal mass, whereas $M_c = 1.2$ GeV and $M_b = 4.75$ GeV) or gluon jet and $m_g(T) =\mu_E(T)/ \sqrt{2}$~\citep{mg} is the effective gluon mass in finite temperature QCD medium. Note that for all parameters we use standard literature values, i.e.,  we do not include additional fitting parameters when comparing our predictions with experimental data.

The analytical expression for the collisional energy loss per unit length in the 
evolving medium is given by~\citep{DREENAB}:
\begin{align}~\label{collEL}
 & \frac{dE_{coll}}{d \tau} =  \frac{2 C_R}{\pi v^2} \alpha_s(ET) \alpha_s(\mu^2_E(T)){\int_0^{\infty}} n_{eq}(|\vec{{\mathbf{k}}}|,T)d |\vec{{\mathbf{k}}}| \nonumber \\
 \times & \Big[\int_0^{|{\vec{{\mathbf{k}}}}|/(1+v)} d |\vec{{\mathbf{q}}}| \int_{-v |\vec{{\mathbf{q}}}|}^{v |\vec{{\mathbf{q}}}|} \omega d \omega  + \int_{|{\vec{{\mathbf{k}}}}|/(1+v)}^{|\vec{{\mathbf{q}}}|_{max}} d |\vec{{\mathbf{q}}}|  \int_{ |\vec{{\mathbf{q}}}| -2|\vec{{\mathbf{k}}}| }^{v |\vec{{\mathbf{q}}}|} \omega d \omega \Big]  \\
\times & \Big[|\Delta_L(q,T)|^2 \frac{(2 |\vec{{\mathbf{k}}}| + \omega)^2 - |\vec{{\mathbf{q}}}|^2}{2} +  |\Delta_T(q,T)|^2 \frac{(|\vec{{\mathbf{q}}}|^2 - \omega^2) ((2 |\vec{{\mathbf{k}}}| + \omega)^2 + |\vec{{\mathbf{q}}}|^2)}{4 |\vec{{\mathbf{q}}}|^4} (v^2 |\vec{{\mathbf{q}}}|^2 - \omega^2)\Big], \nonumber
\end{align}
where $n_{eq}(|\vec{{\mathbf{k}}}|, T) = \frac{N}{e^{|\vec{{\mathbf{k}}}|/T} -1} + \frac{N_f}{e^{|\vec{{\mathbf{k}}}|/T} + 1}$ is the equilibrium momentum distribution ~\citep{BT} comprising gluons, quarks and antiquarks ($N=3$ and $N_f=3$ are the number of colors and flavors, respectively). $k$ is the 4-momentum of the incoming medium parton, $v$ is velocity of the initial jet and $q = (\omega, \vec{{\mathbf{q}}})$ is the 4-momentum of the exchanged gluon.  $|\vec{{\mathbf{q}}}|_{max}$ is provided in Ref.~\citep{DColl}, while $\Delta_T (T)$ and  $\Delta_L(T)$ are effective transverse and longitudinal gluon propagators given by Eqs. (3) and (4) in Ref.~\citep{DREENAB}.

One of the assets of our energy loss formalism is the fact that energy loss explicitly depends on $T$, which makes it naturally suited for examining the QGP properties via implementation of various temperature profiles. In this paper, the temperature dependence on proper time ($\tau$) is taken according to the ideal hydrodynamical 1+1D Bjorken expansion~\citep{Bjorken}  $T(\tau) \sim \sqrt[3]{(\tau_0 / \tau)}$, with thermalization time $\tau_0 = 0.6$ fm~\cite{tau0KH,tau0BMB}. The initial QGP temperature $T_0$ for the chosen centrality bin is determined as described in~\citep{DREENAB}. In this paper, we will concentrate on mid central $30-40\%$ centrality region at 5.02 TeV $Pb+Pb$ at the LHC, which corresponds to $T_0=391$ MeV~\citep{DREENAB}. We however performed the extensive study on all centrality regions (as in~\citep{DREENAB}), and checked that the results/conclusions obtained here are the same irrespectively of the centrality region (results not shown for brevity). The QGP transition temperature is considered to be $T_C \approx 160$~\citep{Temp0}. 

DREENA-B framework is applied for generating predictions for two main high-$p_{\perp}$ observables $-$ $R_{AA}$ and $v_2$. 
 $R_{AA}$ is defined as the ratio of the quenched $A + A$ spectrum to the $p + p$ spectrum, scaled
by the number of binary collisions $N_{bin}$:
\begin{eqnarray}
R_{AA}(p_T)=\dfrac{dN_{AA}/dp_T}{N_{\mathrm{bin}} dN_{pp}/dp_T},
\label{RAA}\
\end{eqnarray}
while for intuitive understanding of the underlying effects  we also use~\citep{DREENAC}: 
\begin{align}~\label{RAA0}
R_{AA}\approx \frac{R^{in}_{AA} + R^{out}_{AA}}{2},
\end{align}
where $R^{in}_{AA}$ and $R^{out}_{AA}$ denote in-plane and out-of-plane nuclear modification factors, respectively. The expression for the high-$p_{\perp}$ elliptic flow~\citep{DREENAC, v2high,vP2} reads:
\begin{eqnarray}
v_{2} \approx \frac{1}{2} \frac{R^{in}_{AA} -R^{out}_{AA}}{R^{in}_{AA} + R^{out}_{AA}}.
\label{v20}
\end{eqnarray}

\section{Results and Discussion}

\begin{figure}
  \includegraphics[width=\linewidth]{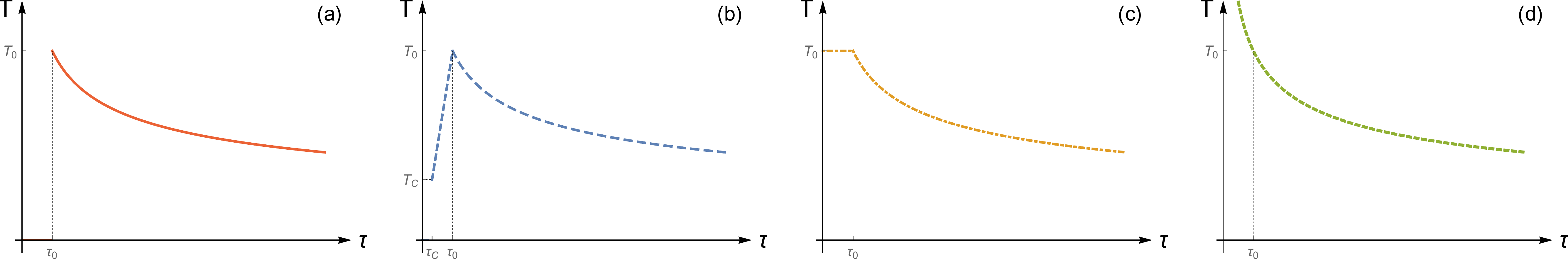}
  \caption{Four temperature evolution profiles, which differ at the initial stages. At $\tau \geq \tau_0$, all profiles assume the same temperature dependence on the proper time (1 + 1D Bjorken~\citep{Bjorken}). At the initial stage, i.e., for $0 < \tau < \tau_0$, the temperature is considered to be: (a) equal to zero; (b) increasing linearly from $T_C$  to $T_0$ between $\tau_C$ and $\tau_0$, otherwise zero; (c) constant and equal to $T_0$; and (d) a continuous function of $\tau$ matching the dependence for $\tau \geq \tau_0$. Note that, in each panel, $T_0$ has the same value at $\tau_0$.}
  \label{fig:sl1}
\end{figure}

In the first part of this section we address how  different initial stages (before the thermalization time $\tau_0$) affect our predictions of high-$p_{\perp}$ $R_{AA}$ and $v_2$. To this end, we consider the following four common cases of initial stages~\citep{SHEME}, which assume the same 1+1D Bjorken hydro temperature ($T$) profile~\citep{Bjorken} upon thermalization (for $\tau \geq \tau_0$), but have different $T$ profiles before the thermalization  (for  $ \tau < \tau_0$):
\begin{enumerate}[(a)]
\item  $T = 0$, the so-called {\it free-streaming case}, which corresponds to neglecting interactions (i.e., energy loss) before the QGP thermalization. 
\item The {\it linear case}, corresponding to linearly increasing $T$ with time from transition temperature ($T_C = 160$ MeV at $\tau_C = 0.25$ fm) to the initial temperature  $T_0$.
\item The {\it constant case} $T = T_0$, and 
\item The {\it divergent case}, corresponding to 1+1D Bjorken expansion from $ \tau=0$. 
\end{enumerate} 
These initial stages are depicted in Fig.~\ref{fig:sl1}, and it is clear that (a)-(d) case ordering corresponds to gradually increasing pre-thermal interactions.  Note that we use this classification (a)-(d) consistently throughout the paper to denote initial stages (for $\tau < \tau_0$), as well as for the entire evolution. Also, note that in this part of the study, we will include experimental data for comparison with our predictions. However, to allow better visualization of our obtained numerical results, in the other two parts of the study  we will omit the comparison with the data, as the error bars are large and the data remain the same. 
 
\begin{figure}
  \includegraphics[width=\linewidth]{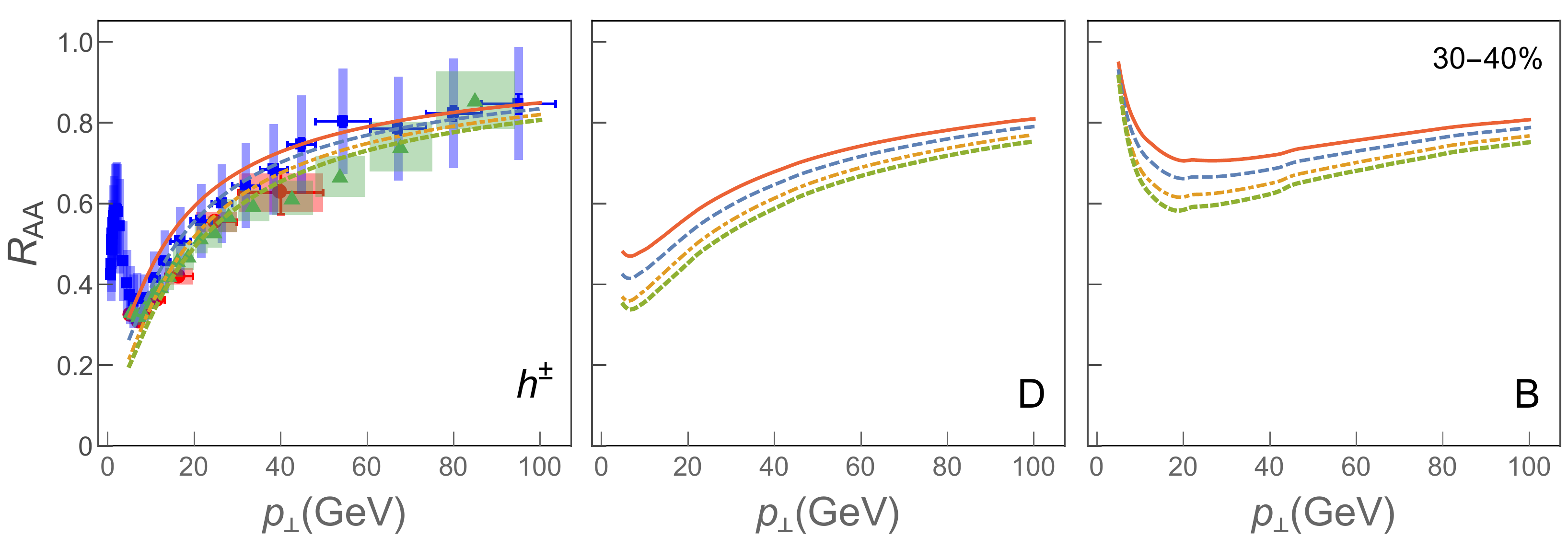}
  \caption{$R_{AA}$ dependence on $p_{\perp}$ for four different initial stages depicted in Fig.~\ref{fig:sl1} is shown for charged particles ({\it left panel}), D mesons ({\it central panel}) and B mesons ({\it right panel}). For charged hadrons, the predictions are compared with $5.02$~TeV $Pb+Pb$ ALICE~\cite{ALICER} (red circles), ATLAS~\cite{ATLASR} (green triangles) and CMS~\cite{CMSR} (blue squares) $h^\pm$ $R_{AA}$ experimental data. In each panel, temperature profile from Fig.~\ref{fig:sl1} are presented by full red curve (case a), by dashed blue curve (case b), by dot-dashed orange curve (case c) and by dotted green curve (case d). The results correspond to the centrality bin $30-40 \%$, and $\mu_M / \mu_E = 0.5$.}
  \label{fig:sl2}
\end{figure}

 Intuitively, one would expect that introducing these pre-thermal interactions would increase the energy loss compared to the commonly considered free-streaming case, and consequently lead to smaller $R_{AA}$. In Fig.~\ref{fig:sl2} we indeed observe that $R_{AA}$ is sensitive to the initial stages. That is, as expected, we see that the suppression progressively increases from case (a) to case (d). However, these differences are not very large, and the current errorbars at the LHC do not allow distinguishing between these scenarios, as can be seen in Fig. 2 (left). 

\begin{figure}
  \includegraphics[width=\linewidth]{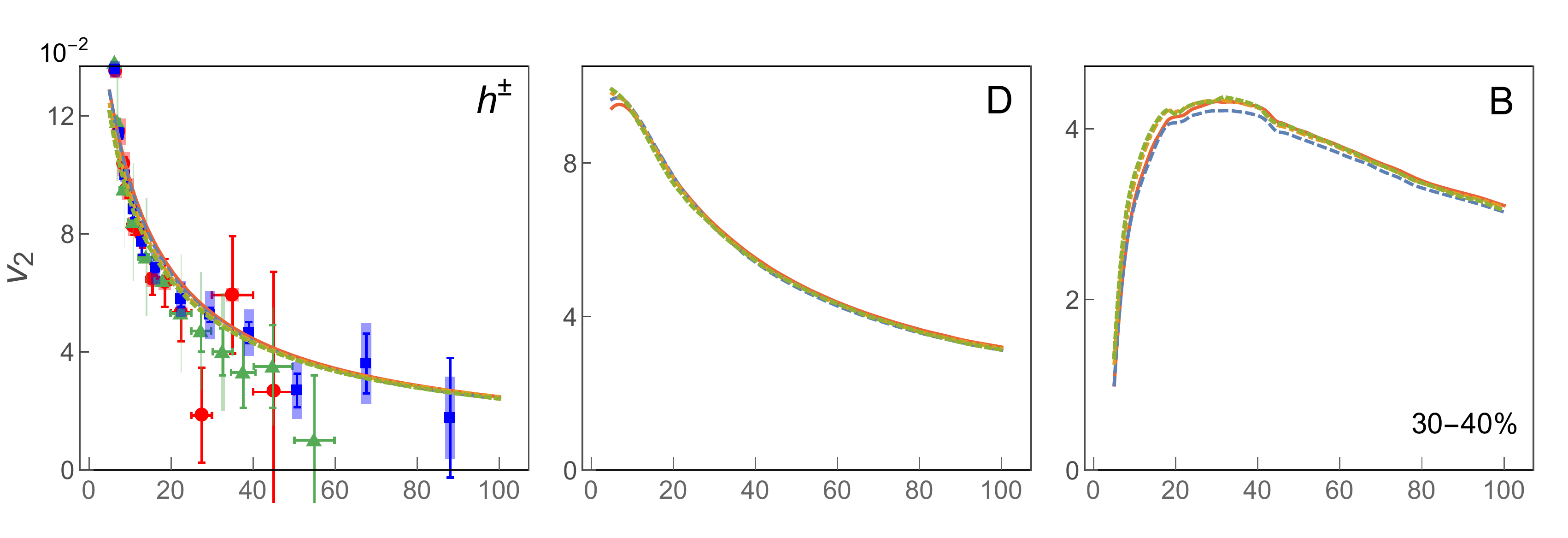}
  \caption{$v_2$ dependence on $p_{\perp}$ for four different initial stages depicted in Fig.~\ref{fig:sl1}. {\it Left}, {\it central} and {\it right panels} correspond to charged hadrons, D mesons and B mesons, respectively. For charged hadrons, the predictions are compared with 30-40\% centrality $5.02$~TeV $Pb+Pb$ ALICE~\cite{ALICEv2} (red circles), ATLAS~\cite{ATLASv2} (green triangles) and CMS~\cite{CMSv2} (blue squares) $h^\pm$ $v_2$ experimental data.  The labeling and remaining parameters are the same as in Fig.~\ref{fig:sl2}.}
  \label{fig:sl3}
\end{figure}

Contrary to $R_{AA}$, the effect of initial stages on $v_2$ is intuitively less clear, as this observable non-trivially depends on the energy loss or $R_{AA}$s (see Eq.~\eqref{v20}). From Fig.~\ref{fig:sl3}, we surprisingly infer that $v_2$ is insensitive to the presumed initial stage  for all types of particles (in distinction to the results obtained in~\citep{Andres}), so that $v_2$ is  unable to distinguish between different initial-stage scenarios.

\begin{figure}
  \includegraphics[width=0.4\linewidth]{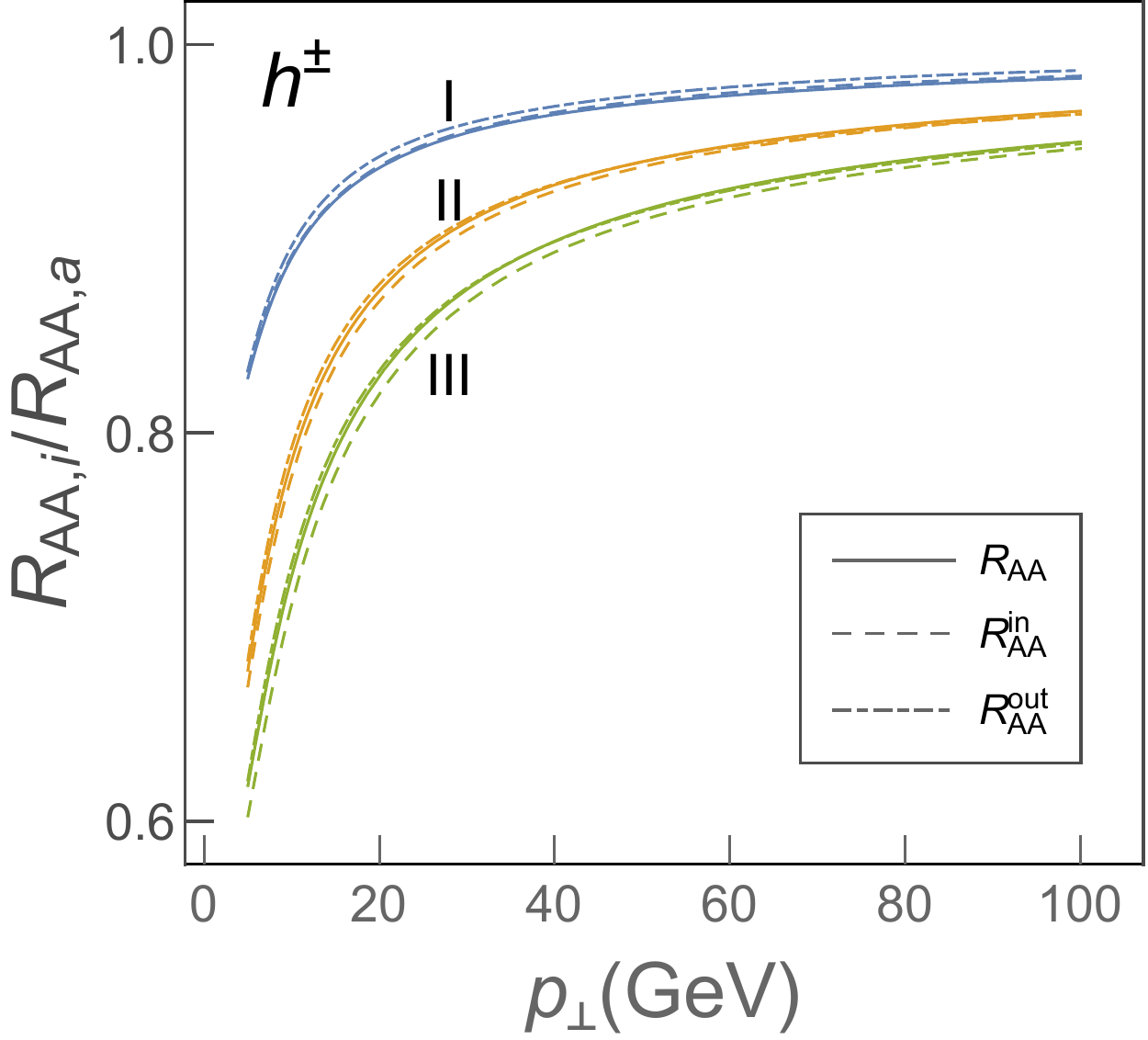}
  \caption{Transverse momentum dependence of in-plane (dashed), out-of plane (dot-dashed) and angular averaged (full curves) $R_{AA}$ relative to the free-streaming case for charged hadrons. Blue (upper), orange (middle) and green (lower) set of curves correspond, respectively, to I, II and III cases. The remaining parameters are the same as in Fig.~\ref{fig:sl2}.}
  \label{fig:sl31}
\end{figure}

To quantitatively understand this unexpected observation, in Fig.~\ref{fig:sl31} we show transverse momentum dependence of $R^{in}_{AA}$, $R^{out}_{AA}$ and $R_{AA}$ in $i=b,c,d$ cases relative to the baseline case (a) for charged hadrons. The conclusions for heavy particles are the same and therefore omitted. We distinguish three sets of curves, which corresponds to the ratio of $R_{AA}$s in a I) linear (b), II) constant (c), III) divergent (d) relative to free-streaming (a) case. Note that the free-streaming case is used as a baseline, as it corresponds to the most commonly used scenario, both in low and high-$p_\perp$ calculations.

Each set of curves in Fig.~\ref{fig:sl31} contains three lines, representing proportionality functions $\gamma(p_{\perp})$s, which are defined as follows:
\begin{eqnarray}
\gamma^{in}_{i a}= \frac{R^{in}_{AA,i}}{R^{in}_{AA,a}},\quad
\gamma^{out}_{i a} = \frac{R^{out}_{AA,i}}{R^{out}_{AA,a}}, \quad
\gamma_{i a} = \frac{R_{AA,i}}{R_{AA,a}},
\label{in}
\end{eqnarray}
where $i =b,c,d$ denotes the corresponding cases from  Fig.~\ref{fig:sl1}. From Fig.~\ref{fig:sl31} we see that for the same $i$ (i.e., within the same set of curves I, II or III) the proportionality functions 
 $\gamma_{i a}(p_{\perp})$ are practically identical for the relations involving in-plane, out-of-plane and angular averaged $R_{AA}$s:
 \begin{eqnarray}
\gamma^{in}_{i a} \approx \gamma^{out}_{i a} \approx \gamma_{i a}.
\label{inout}
\end{eqnarray}
Note also that $\gamma_{i a} < 1$, while $\gamma_{i a}$s from distinct sets significantly differ from one another (i.e., for $i \neq j \rightarrow \gamma_{i a}(p_{\perp}) \neq \gamma_{j a}(p_{\perp})$). 

Consequently, by implementing Eq.~(\ref{in}) in Eq.~\eqref{v20} and acknowledging Eq.~\eqref{inout}, we obtain:  
\begin{eqnarray}
v_{2,i}  \approx \frac{1}{2} \frac{\gamma_{i a} (R^{in}_{AA,a} - R^{out}_{AA,a})}{\gamma_{i a} (R^{in}_{AA,a} + R^{out}_{AA,a})} =v_{2,a},
\label{v2}
\end{eqnarray}
for any choice of $i=b,c,d$, as observed in Fig.~\ref{fig:sl3}. Therefore, we here showed that initial stages alone do not affect $v_2$, i.e., they affect only $R_{AA}$. $R_{AA}$ susceptibility to the initial stages is in a qualitative agreement with papers~\citep{Renk,Molnar,DREENAB}, where $R_{AA}$ is shown to be only sensitive to the averaged properties of the evolving medium, i.e., average temperature ($\overline{T}$). Since $R_{AA}$ is proportional to the $\overline{T}$, and since for all four initial-stage cases (a)-(d) the $\overline{T}$ is different ($\overline{T}_a < \overline{T}_b < \overline{T}_c < \overline{T}_d$), it is evident that $R_{AA}$ will be different in these cases.

\begin{figure}
  \includegraphics[width=0.4\linewidth]{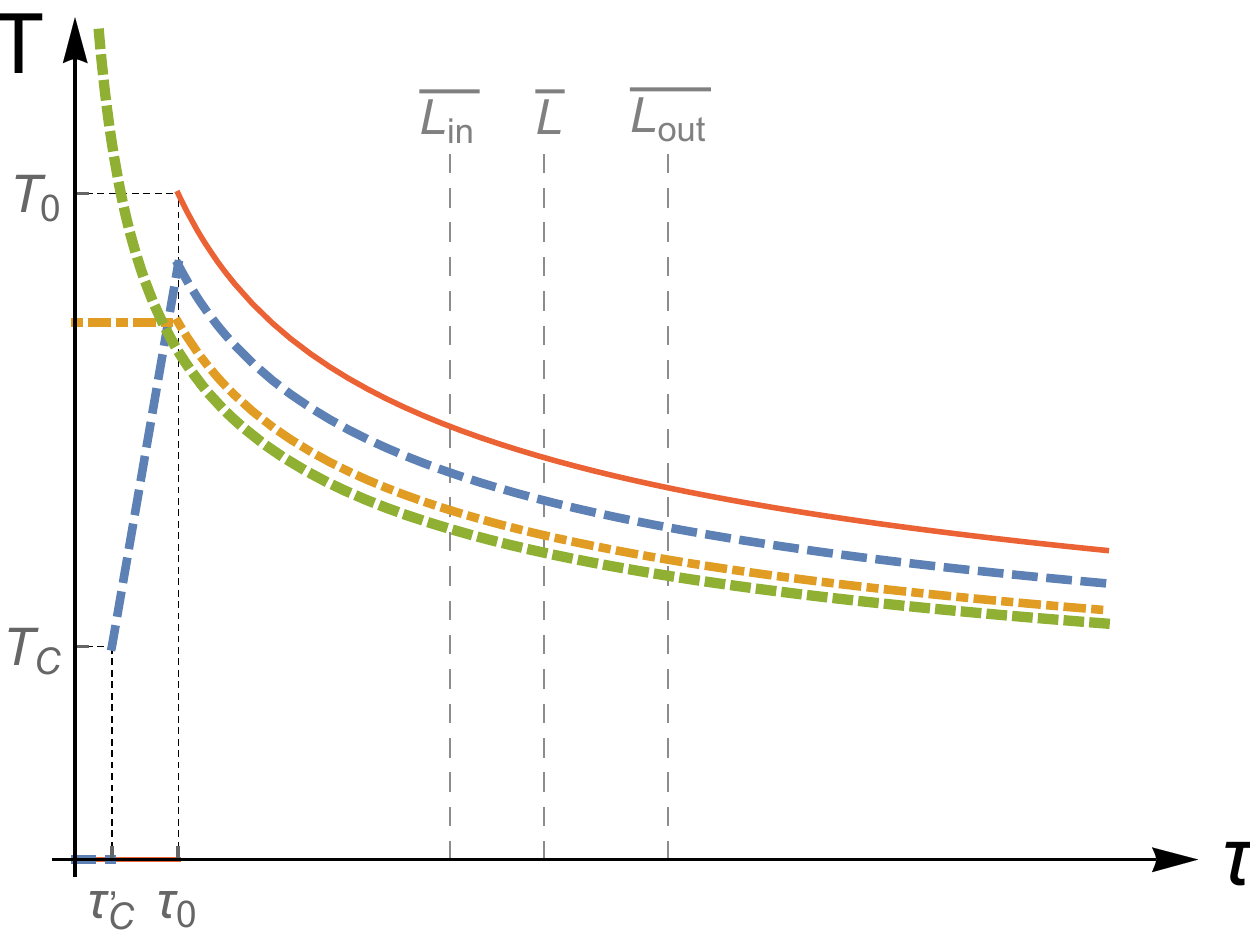}
  \caption{Temperature dependence on the proper time in the setup with the same average temperatures. The labeling is the same as in Fig.~\ref{fig:sl1}, apart from the fact that initial temperatures ($T_0$'s) now differ in these four cases. As in Fig. 1, $T_C = 160$ MeV, $\tau_0 = 0.6$ fm and $\tau'_C = 0.27$ fm. Vertical gray dashed lines correspond to average in-medium path length ($\overline{L}$), and to the path lengths along in-plane ($\overline{L_{in}}$) and out-of-plane ($\overline{L_{out}}$) directions, as labeled in the figure.}
  \label{fig:sl4}
\end{figure}
The fact that $R_{AA}$ depends on the average temperature of the medium, motivate us to further explore the case in which we modify the above temperature profiles to reproduce the same average temperature. This is equivalent to re-evaluating the initial temperatures for different cases from Fig.~\ref{fig:sl1}, and based on the reasoning above, it is evident that new initial temperatures should satisfy the following ordering: $T_{0,d'} < T_{0,c'}< T_{0,b'} < T_{0,a'}$. This leads to $T$ profiles, which do not differ only at early times ($\tau < \tau_0$), but represent {\it different evolutions altogether}. These new evolutions, that are illustrated in Fig.~\ref{fig:sl4} (which is a counterpart of Fig.~\ref{fig:sl1} for the second part of this section), are denoted as (a')-(d') and referred to as "modified" $T$ profiles ((a)$\equiv$ (a')).
\begin{figure}
  \includegraphics[width=\linewidth]{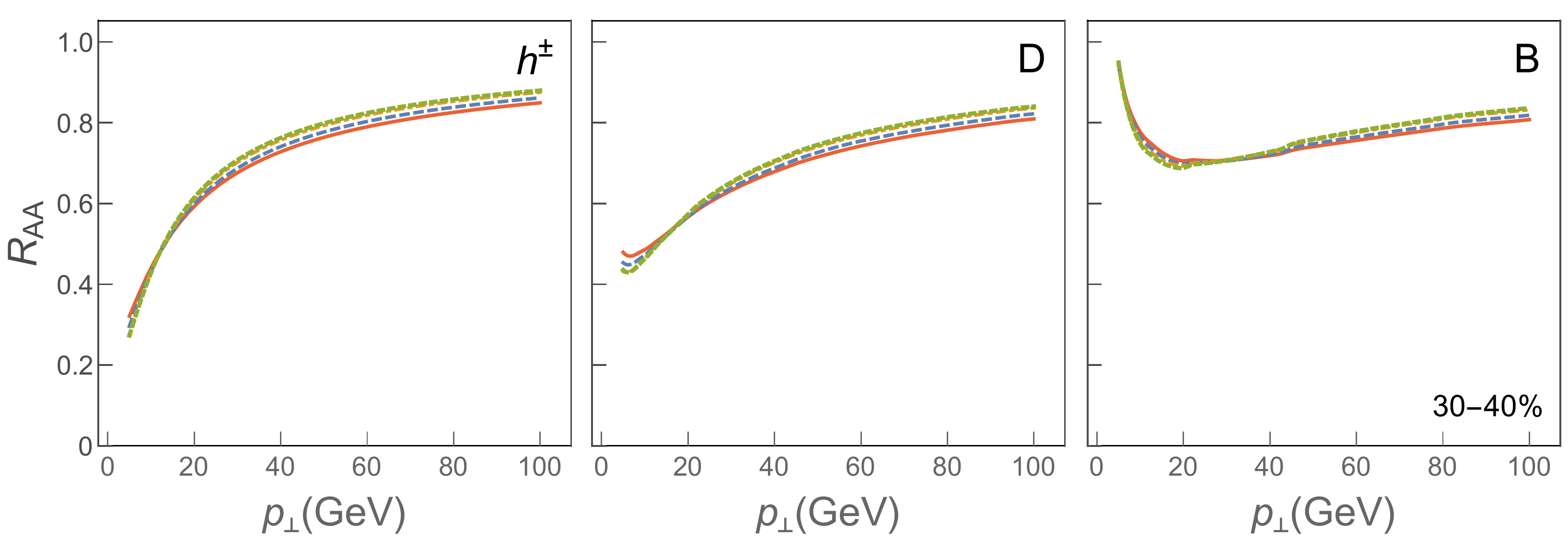}
  \caption{$R_{AA}$ dependence on $p_{\perp}$ for four different medium evolutions depicted in Fig.~\ref{fig:sl4}. {\it Left}, {\it central} and {\it right panels} correspond to charged hadrons, D mesons and B mesons, respectively. In each panel, $T$ profile corresponding to the case: (a') from Fig.~\ref{fig:sl4} is presented by full red curve, (b') dashed blue curve, (c') dot-dashed orange curve and (d') dotted green curve. The results correspond to the centrality bin $30-40 \%$, and $\mu_M / \mu_E = 0.5$.}
  \label{fig:sl5}
\end{figure}

\begin{figure}
  \includegraphics[width=\linewidth]{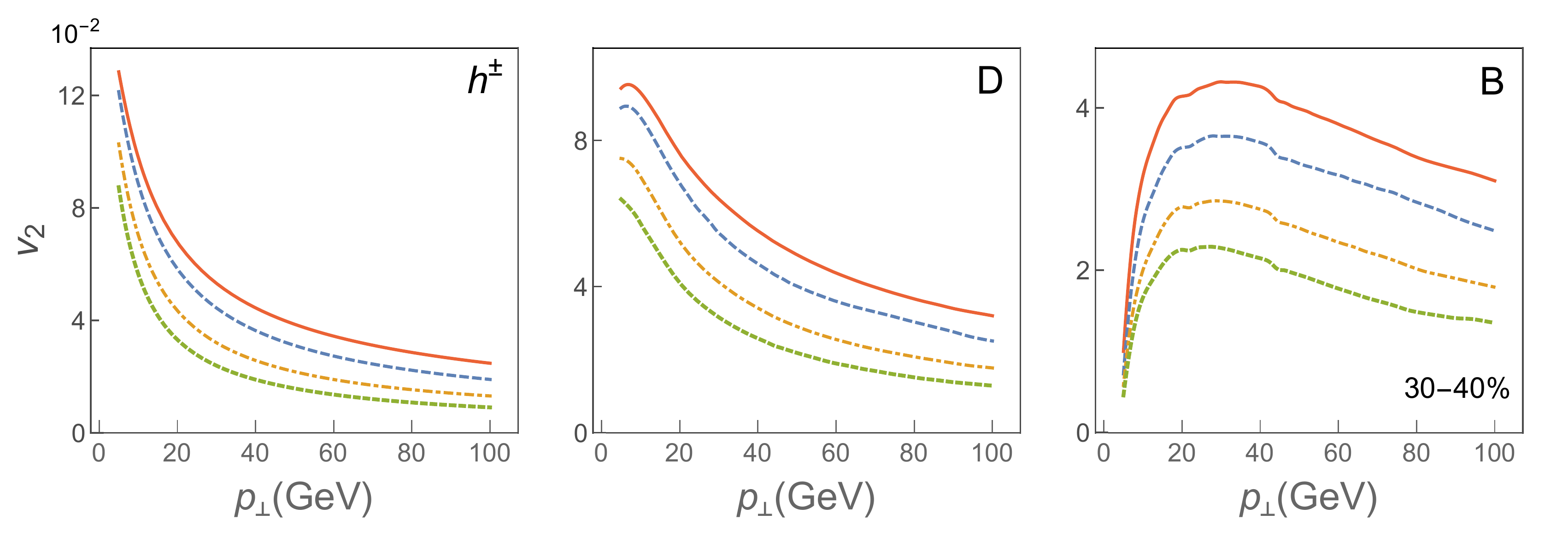}
  \caption{$v_2$ dependence on $p_{\perp}$ for four different medium evolutions depicted in Fig.~\ref{fig:sl4}. {\it Left}, {\it central} and {\it right panels} correspond to charged hadrons, D mesons and B mesons, respectively. The labeling and remaining parameters are the same as in Fig.~\ref{fig:sl5}.}
  \label{fig:sl6}
\end{figure}

In this second $T$-profiles setup, we first verify from Fig.~\ref{fig:sl5} that $R_{AA}$s in all four cases practically overlap, as expected. We next address how these modified evolution cases $(a')-(d')$ affect $v_2$. From Fig.~\ref{fig:sl6} we see that $v_2$ is now very sensitive to the transition from free-streaming case to other modified $T$ profiles. More accurately, for all types of particles, the lowest $v_2$ is observed in modified divergent case, while the highest $v_2$ is observed in the free-streaming case. 

  \begin{figure}
  \includegraphics[width=0.4\linewidth]{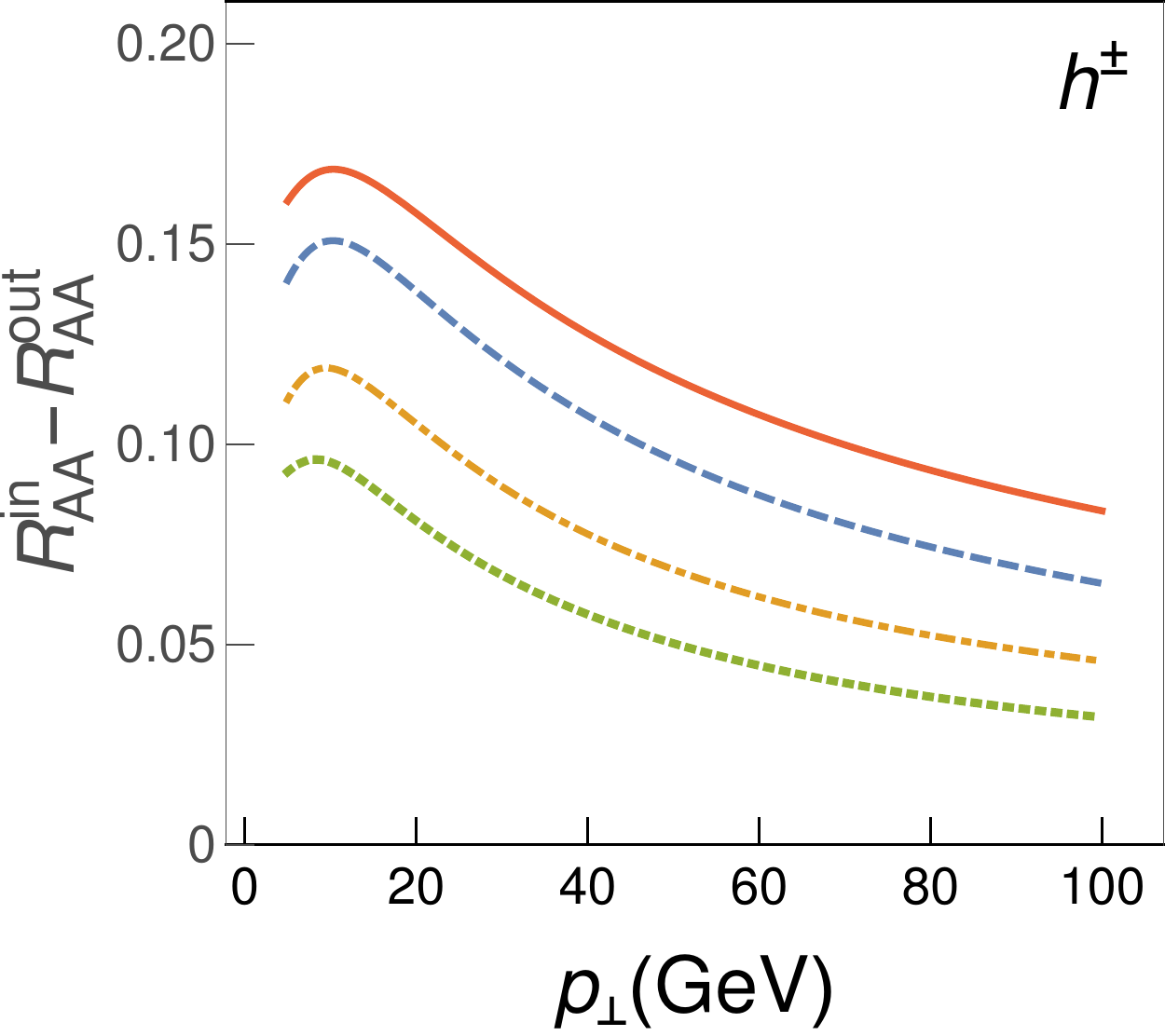}
  \caption{$R^{in}_{AA} - R^{out}_{AA}$ dependence on $p_{\perp}$ for charged hadrons. The labeling and remaining parameters are the same as in Fig.~\ref{fig:sl5}.}
  \label{fig:sl61}
\end{figure} 
The observation from Fig.~\ref{fig:sl6} leads to the following two questions: {\it i)} Why is $v_2$ altered by these modified $T$ profiles $(a')-(d')$? and {\it ii)} Are these discrepancies a consequence of different initial stages? The answer to these questions, we first note that, within this setup, the differences  between $v_2$ (observed in Fig.~\ref{fig:sl6}) are proportional to $R^{in}_{AA} - R^{out}_{AA}$, as the denominator in Eq.~\eqref{v20} (as a starting premise) remains practically unchanged (see Fig.~\ref{fig:sl5}). The transverse momentum dependence of $R^{in}_{AA} - R^{out}_{AA}$ is further shown in Fig.~\ref{fig:sl61} for charged hadrons (as results for D and B mesons will lead to the same conclusion). We see a clear hierarchy, i.e., the largest $R^{in}_{AA} - R^{out}_{AA}$ for free-streaming, descending towards divergent case. To quantitatively understand this observation, we note that for $R_{AA}^{in}$, the high-$p_\perp$ probes traverse, on the average, the medium up to $\overline{L}_{in}$, while for $R_{AA}^{out}$, the medium is traversed up to $\overline{L}_{out}$. Consequently, if we refer to Fig.~\ref{fig:sl4}, $R^{in}_{AA} - R^{out}_{AA}$ comes from $T$-profile difference in the time region between $\overline{L}_{in}$ and $\overline{L}_{out}$, i.e., {\it upon thermalization}. Since in this region $\overline{T}_{d'} < \overline{T}_{c'} < \overline{T}_{b'} < \overline{T}_{a'}$ holds, $R^{in}_{AA} - R^{out}_{AA}$ is the largest for free-streaming case and the smallest for  the divergent case, as observed in Fig.~\ref{fig:sl61}, and in agreement with $v_2$ ordering in Fig.~\ref{fig:sl6}. This therefore provides clarification of why $R^{in}_{AA} - R^{out}_{AA}$, and consequently $v_2$, is affected by these four different QGP evolution profiles, and that this difference originates primarily from the interactions of high-$p_{\perp}$ parton with {\it thermalized QGP}, and {\it not the initial stages}. This agrees with the first part of this section (Figs.~\ref{fig:sl2} and~\ref{fig:sl3}), where we showed and explained insensitivity of $v_2$ to different initial stages. It is worth emphasizing that, contrary to the  first part of this section, in the second part we tested the effects on $R_{AA}$ and $v_2$ not from distinctive initial stages, but instead from four entirely different evolutions of the QCD medium (related by the same global property, i.e., average temperature).

\begin{figure}
  \includegraphics[width=0.8\linewidth]{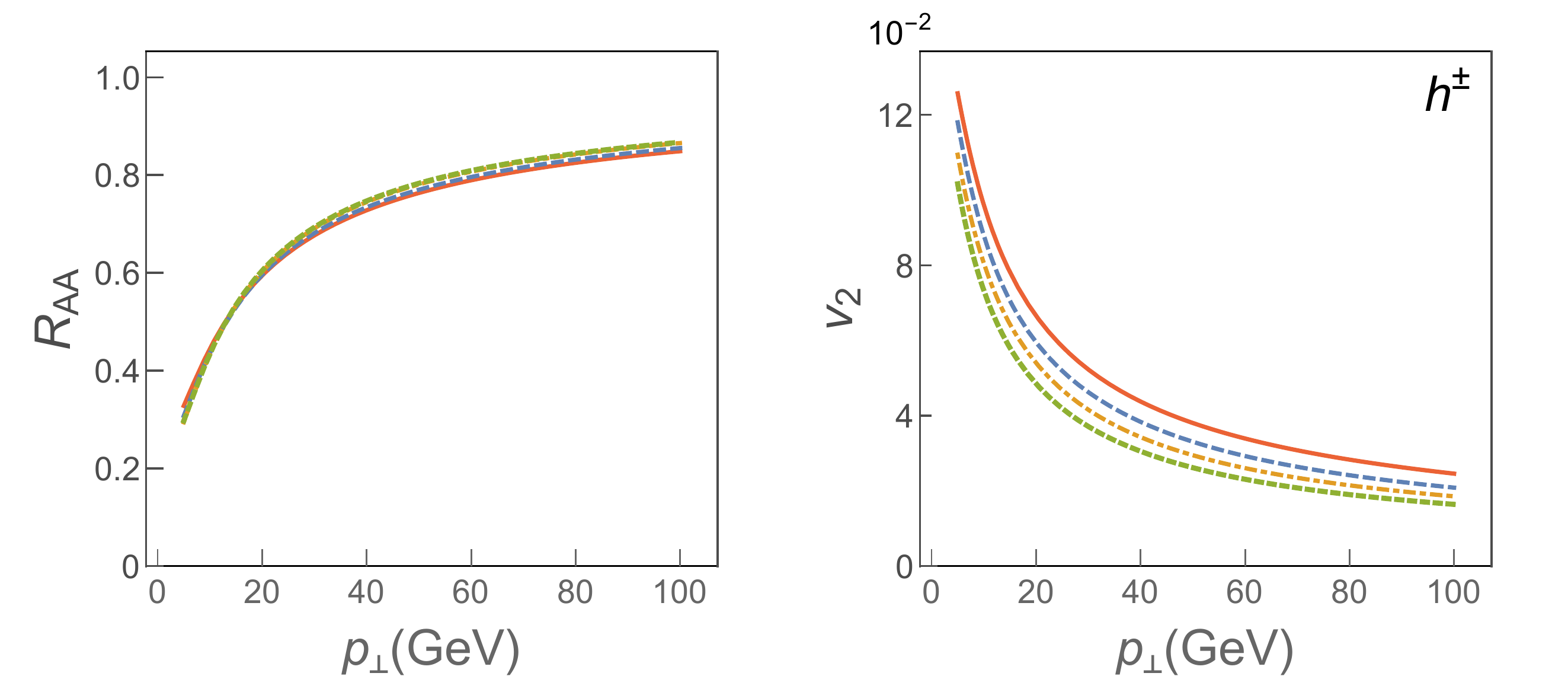}
  \caption{$R_{AA}$ ({\it{left panel}}) and $v_2$ ({\it{right panel}}) dependence on $p_{\perp}$ for charged hadrons, when additional energy loss multiplicative factor is introduced to reproduce the free-streaming $R_{AA}$, in four different initial-stage cases depicted in Fig.~\ref{fig:sl1}. The labeling and remaining parameters are the same as in Figs.~\ref{fig:sl2} and ~\ref{fig:sl3}.} 
  \label{fig:sl9}
\end{figure}
In the final, third, part of this section we adopt a commonly used approach, in which the energy loss is fitted through change of multiplicative fitting factor in the energy loss, to reproduce the desired high-$p_{\perp}$  $R_{AA}$, e.g., the one that best fits the experimental data (see e.g.,~\citep{Andres,vP2,vP3,06,DAB,JETSC}). To this end, we use the same four $T$-profiles from the first part of this section (Fig.~\ref{fig:sl1}), while, in our full-fledged calculations (see Sec. II) we introduce an additional multiplicative fitting factor (free parameter) $C^{fit}_i$, $i=b,c,d$. $C^{fit}_i$ is then estimated for each initial-stage case as a best fit to the free-streaming $R_{AA}$ (see Table~\ref{table2}). Thus obtained $R_{AA}$s are shown in the left panel of Fig.~\ref{fig:sl9} only for the representative case of $h^{\pm}$, as the same conclusions stand for both light and heavy flavor hadrons. From the left panel of this figure we observe practically overlapping $R_{AA}$s in all (a)-(d) cases, as anticipated, which is obtained by decreasing $C^{fit}_i$ consistently from the free-streaming to the divergent case (each $C^{fit}_i \leq 1$) in order to {\it compensate} for the higher energy losses in the corresponding cases compared to the case (a).

\begin{table}[ht]
\centering
\begin{tabular}{c c}
\hline
\hline
$T$ profile case & $C^{fit}_i$ \\
\hline
Free-streaming case (a) & 1  \\
Linear case (b) & 0.87 \\ 
Constant case (c) &  0.74 \\
Divergent case (d) & 0.67 \\
\hline
\hline
\end{tabular}
\caption{Fitting factors values}
\label{table2}
\end{table}

The effect of different $T$-profiles from Fig.~\ref{fig:sl1} after introduction of multiplicative fitting factor $C^{fit}_i$ in full-fledged numerical procedure on $v_2$ is depicted on the right panel of  Fig.~\ref{fig:sl9}, where we see that elliptic flow in (a)-(d) cases notably differs, i.e., is the highest in the free-streaming case, while the lowest in the divergent case. Based on this observation, one could naively infer that initial stages, i.e., $\tau < \tau_0$ region (the only region in which $T$ profiles differ), have a significant effect on $v_2$, as recently observed by alternative approach~\citep{Andres}. 

However, this kind of reasoning is inconsistent with our analysis outlined in the first two parts of this section, as well as with intuitive expectation that introduction of the energy loss at the  initial stage affects $R_{AA}$. To quantitatively understand this result, we introduce asymptotic scaling behavior~\citep{DREENAC,DREENAB,NewObserv}. That is, for higher $p_\perp$ of the initial jet, and for higher centralities (where fractional energy loss is expected to be small), we can make the following estimates: 
\begin{eqnarray}
\Delta E/E &\approx& \chi \overline{T}^m \overline{L}^n, \\
R_{AA} &\approx& 1-\frac{l-2}{2} \frac{\Delta E}{E} = 1-\xi \overline{T}^m \overline{L}^n
\label{est}
\end{eqnarray}
where $m, n$ are proportionality factors, $\overline{T}$ is the average temperature of the QGP, $\overline{L}$ denotes the average path length traversed by the jet, $\chi$ is a proportionality factor (that depends on $p_\perp$ and flavor of the jet). $\xi=\frac{l-2}{2} \chi$, where $l$ is the steepness of a power law
fit to the transverse momentum distribution.

If $\Delta E/E$ is fitted by additional multiplicative factor $C$, the new $R^{fit}_{AA}$ becomes
\begin{eqnarray}
R^{fit}_{AA,i} \approx 1-C_{i} \xi \overline{T}^m_{i} \overline{L}^n_{i} \approx 1-C_{i} (1- R_{AA, i}),
\label{R31}
\end{eqnarray}
where $i= b,c,d$ and $C_i$ ($C_i < 1, \forall {i}$) denotes the fitting factor, and the last part of Eq.~(\ref{R31}) is obtained by using Eq.~(\ref{est}), leading to
\begin{eqnarray}
C_{i} \approx \frac{1-R^{fit}_{AA,i}} {1- R_{AA, i}},
\label{R311}
\end{eqnarray}
We note that Eq.~(\ref{R311}) is applicable to  the average, in-plane and out-of-plane $R_{AA}$s, since the same fitting factor is consistently applied in all three cases. By imposing the condition (which quantifies the equivalence of fitted $R_{AA}$ in (b)-(d) cases to the free-streaming case):
\begin{eqnarray}
R^{fit}_{AA,i} = R_{AA,a},
\label{R32}
\end{eqnarray}
and by applying  Eqs.~(\ref{RAA0})-(\ref{inout}) and (\ref{R32}), together with Eqs.~(\ref{est},~\ref{R31}) and their in-plane and out-of-plane analogons, we obtain:
\begin{eqnarray}
v^{fit}_{2,i} \approx  \frac{1}{2} \frac{C_{i} (R^{in}_{AA,i} - R^{out}_{AA,i})}{2 R_{AA,a}} = \frac{1}{2} \frac{C_{i} \gamma_{i}(R^{in}_{AA,a} - R^{out}_{AA,a})}{R^{in}_{AA,a} + R^{out}_{AA,a}} = C_{i} \gamma_{i a} v_{2,a},
\label{v3}
\end{eqnarray} 
which can also be written as
\begin{eqnarray}
C_{i} \approx \frac{v^{fit}_{2,i}}{\gamma_{i a} v_{2,a}}.
\label{v31}
\end{eqnarray} 
From Eq.~\eqref{v3}, we see that decrease of $v^{fit}_2$ in (b)-(d) cases compared to (a) is a result of a fitting factor $C_i(p_{\perp})$ (which is smaller than 1), as well as the proportionality functions $\gamma_i(p_{\perp})$ (also smaller than 1).   However, note that Eq.~\eqref{v3} describes asymptotic behavior at very high $p_{\perp}$, where, as mentioned earlier, $\gamma$s approach 1. Consequently, the diminishing of elliptic flow compared to the case (a) is predominantly due to a decrease of the {\it artificially imposed fitting factor} $C$. Therefore, we obtain that, contrary to~\citep{Andres}, {\it initial stages are not} mainly responsible for the obtained differences (the right panel of Fig.~\ref{fig:sl9}) in the $v^{fit}_2$ curves for different $T$ profiles. Moreover, this argument, as well as the obtained inconsistency of the results in this and the first two parts of the paper, implies that application of multiple fitting procedure for each different initial stage may result in incorrect energy loss estimates and in misinterpreting the underlying physics. 

\begin{figure}
  \includegraphics[width=\linewidth]{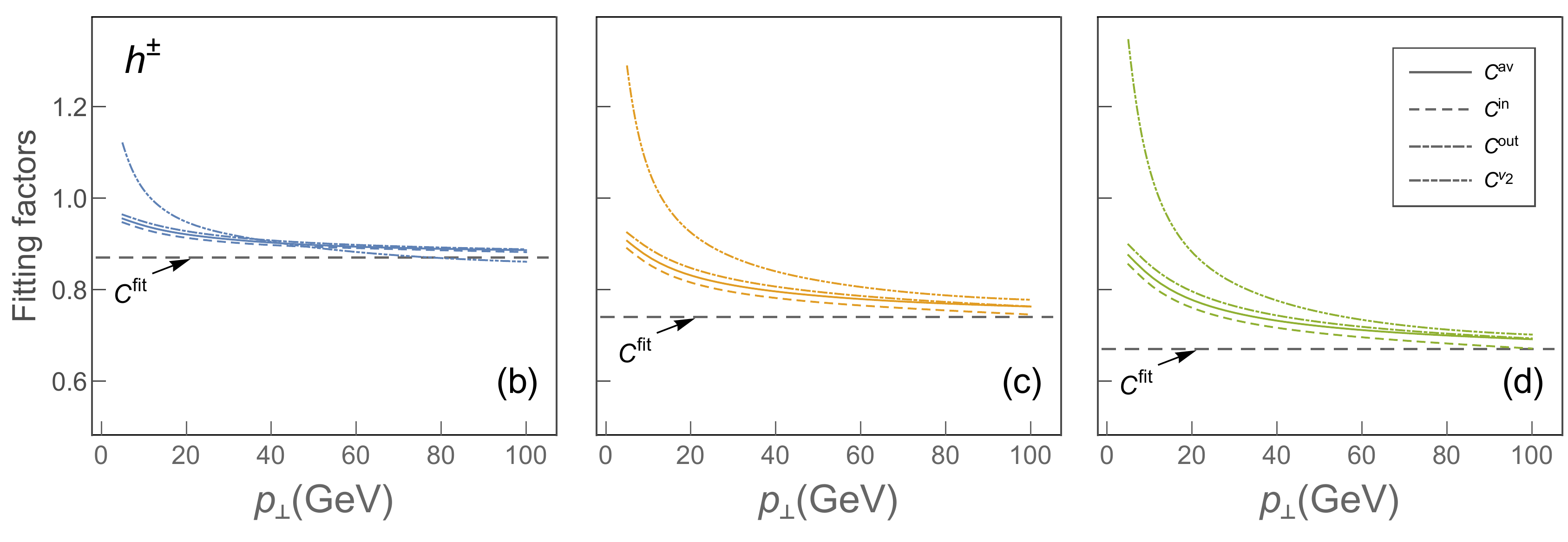}
  \caption{Comparison of four fitting factors defined by Eq.~\eqref{Cs} with $C^{fit}_i$ value, obtained from full-fledged numerical procedure, in {\it{linear}} (b) (left), {\it{constant}} (c) (central) and  {\it{divergent}} (d) (right panel) cases. C factors presented  by full, long dashed, dot-dashed and dot-dot-dashed curves correspond to $h^{\pm}$ angular averaged, in-plane, out-of-plane $R_{AA}$ and $v_2$ cases, respectively. The horizontal gray dashed line presents energy loss fitted value $C^{fit}_i$. The results correspond to the centrality bin $30-40 \%$, and $\mu_M / \mu_E = 0.5$.}
  \label{fig:sl10}
\end{figure}
To asses if this qualitative conclusion indeed holds, i.e. that $v_2$ succesibility observed in Fig.~\ref{fig:sl9} (as well as in~\cite{Andres}) is indeed a consequence of a fitting factor in the energy loss, in Fig.~\ref{fig:sl10} we check the consistency of Eqs.~(\ref{R311}) and~(\ref{v31}) with the full-fledged numerical calculations. That is, a non-trivial consequence of Eqs.~(\ref{R311}) and~(\ref{v31}), is that $C_i$ factors for the average, in-plane and out-of-plane $R_{AA}$s (Eq.~\ref{R311}) and $v_2$ (Eq.~\ref{v31}), should be the same in high-$p_\perp$ limit, and moreover overlap with $C^{fit}_i$ in this limit. To this end, we define the following $C$ factors (originating from Eqs.~(\ref{R311},~\ref{v31})):
\begin{align}~\label{Cs}
& C^{in}_i = \frac{1-R^{in,fit}_{AA,i}}{1-R^{in}_{AA,i}},\quad
C^{out}_i = \frac{1-R^{out,fit}_{AA,i}}{1-R^{out}_{AA,i}} , \nonumber \\
& C^{av}_i = \frac{1-R^{fit}_{AA,i}}{1-R_{AA,i}} ,\quad {C}^{v_2}_i = \frac{1}{\gamma_{ia}} \frac{v^{fit}_{2,i}}{v_{2,a}},
\end{align}
and compare them with $C^{fit}_i$, for each separate initial-stages case, $i=b,c,d$. Note that, while expression themselves on the right-hand side of each expression in Eq.~(\ref{Cs}) are obtained in high-$p_\perp$ limit (and consequently are expected to overlap in this limit, if our analytical estimate is valid), we calculate $C^{fit}_i$, and the terms on the the right-hand side of each expression in Eq.~(\ref{Cs}), through full-fledged numerical procedure.
We indeed observe that, for each $i$ and at high-$p_\perp$: $C^{in}_i$, $C^{out}_i$, $C^{av}_i$ and  $C^{v_2}_i$ factors are practically overlapping, and approach the value $C^{fit}_i$. Consequently, this highly non-trivial observation confirms that our qualitative conclusion is valid, and that $v_2$ susceptibility in this case is indeed a consequence of an additionally introduced fitting factor.

\section{Conclusions} 
Traditionally, the features of initial stages before QGP thermalization are explored through comparison of bulk medium simulations and low-$p_{\perp}$ data. On the other hand, recent abundance of high-$p_{\perp}$ experimental data, motivates exploiting the high-$p_{\perp}$ energy loss in studying the initial stages. We here utilized state-of-the-art dynamical energy loss embedded in analytical 1+1D Bjorken medium expansion (DREENA-B framework), which allowed to tightly control the analyzed temperature profiles. In particular, we considered four temperature profiles, which are identical after, but are different before, thermalization, which correspond to four commonly considered initial-stage cases. This allowed to study the effects of different initial-stage cases on high-$p_\perp$ $R_{AA}$ and $v_2$ predictions, under highly controlled conditions, by combining full-fledged numerical results and analytical estimates used to interpret the experimental results. 

We found that high-$p_\perp$ $R_{AA}$ is sensitive to the pretermalized stages of the medium evolution, however, within the current errorbars, the senistivity is not sufficient to distinguish between different scenarios. On the other hand, the high-$p_\perp$ $v_2$ is unexpectedly insensitive to the initial stages. We furthermore found that previously reported sensitivity~\cite{Andres} of high-$p_\perp$ $v_{2}$ to initial stages is mainly a consequence of the fitting procedure in which the parameters in the energy loss are adjusted to reproduce experimentally observed $R_{AA}$, individually for different initial-stage cases. On the other hand, if the same global property, in particular the same average temperature, is imposed to tested temperature profiles, high sensitivity of high-$p_\perp$ $v_2$ is again obtained. This sensitivity is, however, a consequence of differences in final, rather than initial, stages. Overall, our results underscore that the simultaneous study of high-$p_\perp$ $R_{AA}$ and $v_2$, with consistent/fixed energy loss parameters across the entire study and controlled temperature profiles (reflecting only the differences in the initial stages), is crucial to impose accurate constraints on the initial stages.

{\em Acknowledgments:} 
We thank Pasi Huovinen and Jussi Auvinen for useful discussions. This work is supported by the European Research Council, grant ERC-2016-COG: 725741, and by the Ministry of Science and Technological
Development of the Republic of Serbia, under project No.  ON171004 and ON173052.
 
\end{document}